\documentclass[11pt]{article}
\usepackage{geometry}
\usepackage{xcolor}
\usepackage{a4}
\usepackage{graphicx}
\usepackage{epsf}
\usepackage{amsmath}
\usepackage{amssymb}
\usepackage{braket}
\usepackage{cite}
\newcommand{\be}{\begin{equation}}
\newcommand{\ee}{\end{equation}}

\newcommand{\Rmnum}[1]{\expandafter\@slowromancap\romannumeral #1@}
\newcommand{\bea}{\begin{eqnarray}}
\newcommand{\eea}{\end{eqnarray}}

\begin{document}
\def\A{{\mathbb{A}}}
\def\B{{\mathbb{B}}}
\def\C{{\mathbb{C}}}
\def\R{{\mathbb{R}}}
\def\s{{\mathbb{S}}}
\def\T{{\mathbb{T}}}
\def\Z{{\mathbb{Z}}}
\def\W{{\mathbb{W}}}
\begin{titlepage}
\title{Geometry of AdS black hole thermodynamics in extended phase space}
\author{}
\date{
Joy Das Bairagya, Kunal Pal, Kuntal Pal, Tapobrata Sarkar
\thanks{\noindent E-mail:~ joydas, kunalpal, kuntal, tapo @iitk.ac.in}
\vskip0.4cm
{\sl Department of Physics, \\
Indian Institute of Technology,\\
Kanpur 208016, \\
India}}
\maketitle
\abstract{We consider the geometry of anti-de-Sitter (AdS) black hole thermodynamics in four dimensions, 
where the equation of state in the extended phase space formalism allows explicit comparison with 
normal fluid systems. We show that for the two-dimensional parameter manifolds
considered here, the scalar curvature is proportional to the thermodynamic volume. This allows us to critically 
examine the applicability of geometric methods in black hole thermodynamics in extended phase space. 
We show how several standard features that are expected to hold in normal fluid systems impose severe restrictions
on the black hole parameters, leading to the fact that several results in the current literature on the geometry of
thermodynamics in extended phase space may be physically invalid. These are true for both charged and rotating AdS black holes. As
a by-product of our analysis, we examine a conjecture regarding the equality of the correlation lengths
of co-existing phases near criticality, in charged AdS black hole backgrounds, and find reasonable validity.}
\end{titlepage}

\section{Introduction}

In the absence of a fully understood quantum theory of gravity, coarse-grained approaches are often used
to elucidate broad features of the physics of black holes. Understanding the 
thermodynamic properties of black holes is one such area of research, 
that is immensely popular in the current literature, with the hope being that the results obtained herein should be finally 
related to the micro structure of black holes derived in a more satisfactory quantum formalism. Black hole thermodynamics, which
was formulated several decades ago, is based on the fact that the four laws of black hole mechanics are formally
identical to the four laws of thermodynamics, with the black hole being an equilibrium thermal state with a
non-zero Hawking temperature. 

Ordinary thermodynamic systems on the other hand are by now known to be amenable to a Riemannian 
geometric treatment. The basic idea here is that the tunable parameters of such systems (for example the 
temperature, pressure etc.) 
form a non-trivial parameter manifold, whose geometry captures physical thermodynamic features, the most important
aspect being phase transitions. In the geometric approach, the line element on the parameter manifold 
(which, for the purpose of the discussion here, will always be two-dimensional) 
is indicative of the distance between two thermodynamic states that are related by a fluctuation. Further, 
for systems that exhibit continuous second order phase transitions, the scalar curvature (i.e the Ricci scalar) 
on the parameter manifold is conjectured to be related to the correlation length of the system, and diverges 
near criticality. This approach (popularly known as thermodynamic geometry) has a long history (see e.g.,
the review \cite{BrodyHook}) and was formalized by Ruppeiner \cite{Rupp} (for a pedagogical exposition, 
see \cite{RuppAJP}) for thermodynamic systems, 
and provides a useful alternative method to study thermodynamics of fluids. It is indeed satisfying that, as is well known 
and as we will discuss in details, the geometric formalism inherits its own limits of applicability, due to some 
inherent physical features. 

A natural question that arises in this context is the applicability of Riemannian geometric methods in black hole thermodynamics. 
This becomes particularly interesting in extended phase space thermodynamics, in theories that
admit a cosmological constant. Here, following the proposal of \cite{Ray}, one identifies the variable cosmological
constant as the pressure of the black hole system, with the conjugate volume being the thermodynamic volume, 
and the mass of the black hole is identified with its enthalpy. Once the formal notions
of pressure and volume are established for a black hole, one is tempted to think that geometric methods
applicable to fluids should be of relevance here, to glean insight into black hole thermodynamics in extended phase space. 
And since in ordinary fluid systems, there are stringent restrictions on the applicability of geometric methods
as we have mentioned, it is natural to ask how far these restrict
the applications of such geometric methods to black hole systems. This is the question we ask in this paper. 

Here, one needs to be careful at the outset, given the fact that one of the basic 
premises of thermodynamic geometry, namely the extensivity of the entropy, is invalid for black holes. 
As is standard in the literature however, in constructing the geometry of black hole thermodynamics, 
one proceeds to construct the line element exactly in the same fashion as is done for usual thermodynamic systems.
And, as is well known by now, thermodynamic geometry of black hole systems often exhibit the same
behaviour as usual fluid systems, thus providing useful insights into the coarse grained nature of 
black hole microstates (see, e.g., the earlier works by one of the present authors \cite{TSS} - \cite{SSS3} as well as
the works of \cite{New1} - \cite{New3}, and references therein). 

In the era before the notion of extended phase space was introduced, thermodynamic geometry of various
classes of black holes were worked out in scenarios that did not involve a pressure or volume. These were 
studied in various ensembles that typically include two fluctuating parameters, often taken as the mass
and a charge, or the mass and the corresponding potential. For charged anti-de-Sitter (AdS) black holes, 
the pioneering works of \cite{Chamblin1},\cite{Chamblin2} established liquid-gas like phase transition
properties, with an equation of state specifying the temperature as a function of the charge and the electric potential. 
Methods of thermodynamic geometry were used soon after these works, 
to understand such phase transitions from a geometric perspective, and useful scaling relations involving
the black hole parameters were obtained in several works. In extended phase space however, as we will discuss in detail, 
one has on the other hand an explicit equation of state involving the black hole pressure and volume, and hence 
geometric notions typical to fluid systems are more directly applicable here. 

Indeed, one of the main lessons from the study of thermodynamic geometry in fluids is the relationship between the 
scalar curvature $R$ and the correlation length $\xi$, via $R \sim \xi^d$ near criticality, with
$d$ being the spatial dimension of the system, as can be shown via a renormalization group approach \cite{tapo2}. 
For two dimensional parameter manifolds that we will be interested in, the Riemann curvature tensor has
a single independent component, and hence $R$, which fully characterizes the curvature of the manifold,
plays a fundamental role. At the very outset, the relationship between $R$ and the correlation volume entails a dimensional
constraint on $R$, namely that $R$ {\it must} scale as a volume dimension. This in fact is crucial in 
determining the applicability of the geometric method itself. Namely, in a coarse-grained approach 
(for example in a fluid system),
the geometric analysis loses meaning if $R$ is of the order of, or less than, a typical molecular volume \cite{RSSS}. 
Also, by its very construction, the geometry assumes a Gaussian approximation to the 
expansion of the entropy around an equilibrium value, and it is important to understand if the validity of this
approximation is ensured in suitable regions of the parameter space. 
All these assume importance in black hole thermodynamics in extended phase space, with a well defined notion of 
thermodynamic volume as well as specific volume, i.e., volume per molecule \cite{AKMS},\cite{WeiLiu}. 

In this paper, we will examine these issues, with our focus being four dimensional charged black holes
as well as rotating black holes
in anti-de-Sitter space, called Reissner-Nordstrom-AdS (RN-AdS) and Kerr-AdS (KAdS) black holes, respectively.
One of the first difficulties that one encounters here is the fact that for RN-AdS black holes, the (dimensionless) 
specific heat capacity at constant volume $c_v=0$ identically, while it vanishes in the slow rotation approximation for
Kerr-AdS black holes. This renders the line element, one of whose components is proportional to $c_v$
(see Eq. (\ref{lineoriginal}) in the next section), meaningless at first sight. An important idea in this context was put
forward in the work of \cite{WeiLiuMannPRL}. These authors suggested that one should generically consider 
the limit $c_v\to 0$ in the final expression of the scalar curvature of the geometry, instead of beginning from it. In this approach,
one can define a normalized scalar curvature $Rc_v$ which diverges at criticality, is finite away from it,
and can be related to the correlation volume near criticality. $Rc_v$ is then the scalar curvature
appropriate for extended phase space thermodynamics of RN-AdS or Kerr-AdS black holes. 
Since $Rc_v$ scales as the correlation
volume, and is the only scalar quantity in two dimensional parameter spaces that we consider, the 
inherited properties of this quantity naturally determines the limits of applicability of the Riemannian geometric method
in theories with vanishing specific heat. 

In this work, we show that $Rc_v$ is a natural physical quantity in closed thermodynamic fluid systems,
where the volume is treated as a fluctuating variable, and establish its relation with the corresponding
quantity in open fluid systems. These are shown to be proportional, with the system volume being the proportionality constant,
in the limit $c_v \to 0$. 
This simple observation then allows us to compare several features of the geometry of thermodynamics 
of standard fluids to that of AdS black holes. In particular, we examine the applicability of geometric 
methods in black hole systems, both for RN-AdS and Kerr-AdS black holes, and
show why many of the results that have recently appeared in the literature might in fact be invalid. 
Further, as a by-product of our analysis, we examine a recent conjecture \cite{RSSS} that the correlation
lengths are equal in coexisting phases, near criticality, and find reasonable validity, in the context
of RN-AdS black holes. This paper is
organized as follows. In section \ref{TG}, we review some of the basic features of thermodynamic geometry. 
In section \ref{OC}, we establish the relation between the scalar curvatures of open and closed thermodynamic fluid systems
in the limit of vanishing specific heat. 
Section \ref{RNAdS} addresses the relevant issues related to the applicability of geometric methods 
to RN-AdS black hole systems, and we
briefly comment on the geometry of Kerr-AdS black holes in section \ref{KAdS}. The paper
ends with a summary and discussions in section \ref{Conclusions}. 

\section{Thermodynamic geometry}
\label{TG}

Classical thermodynamic geometry has its origin in fluctuation
theory. As worked out in the book \cite{LandauLifshitz} by Landau and Lifshitz (section 114), the fluctuation 
probability $w$ in a closed thermodynamic system, with the independent variables taken as the temperature 
$T$ and the volume $V$ can be written as\footnote{The
subscript (or superscript) $c$ will be used to denote geometric quantities in a closed subsystem, 
as opposed to an open subsystem, where we will ue the subscript or superscript $o$.}
\begin{equation}
w \sim {\rm exp}\left[-\frac{1}{2}dl_c^2\right]~,~~
dl_c^2=\frac{C_V}{k_BT^2}dT^2 - \frac{1}{k_BT}\left(\frac{\partial P}{\partial V}
\right)_T dV^2~,
\label{lineoriginal}
\end{equation}
where $k_B$ is Boltzmann's constant, and $P$ denotes the pressure. 
Here, $C_V = T(\partial S/\partial T)_V$ is the specific heat at constant volume. 
This is also written in terms of the dimensionless specific heat $c_v$ as $C_V = Nk_B c_v$, with $N$ being
the number of particles. An early work of Ruppeiner \cite{RuppOriginal} in fact uses the 
$dl_c^2$ of Eq. (\ref{lineoriginal}) as the line element on the 
parameter manifold (with coordinates $T$ and $V$) to show that the scalar curvature $R_c$ 
arising out of this line element diverges 
near the second order critical point in fluid systems.\footnote{That this will be the case for fluid systems that
show a coexistence line culminating in a second order critical point can be seen by noting that the second
term in Eq. (\ref{lineoriginal}) is the inverse of the isothermal compressibility and vanishes along the
spinodal curve. Hence the determinant of the metric of Eq. (\ref{lineoriginal}) is undefined along this curve and
in particular at the critical point.} The exponent in Eq. (\ref{lineoriginal}) and hence $dl_c^2$ 
being naturally dimensionless, $R_c$
also inherits the same property, and \cite{RuppOriginal} considers the quantity $R_cV$ 
and relates it to the correlation volume of the system near criticality. 

Since this is somewhat ad-hoc, a more methodical approach elaborated upon in \cite{Rupp} and in
subsequent literature, is to consider the line element obtained from an appropriate thermodynamic
potential per unit volume, in an open thermodynamic system with a fixed volume $V$. In this case,
it is more convenient to consider the temperature and the density $\rho = N/V = 1/v$ as the fluctuating 
variables. The line element for such an open thermodynamic system is given, for a 
single component fluid, by \cite{Rupp}
\begin{equation}
dl_o^2 = \frac{1}{k_BT}\left(\frac{\partial s}{\partial T}\right)_{\rho} dT^2 + \frac{1}{k_BT}
\left(\frac{\partial \mu}{\partial\rho}\right)_Td\rho^2~,
\label{lineperuv}
\end{equation}
where $s = S/V= -(\partial f/\partial T)_{\rho}$ and $\mu = (\partial f/\partial \rho)_T$ being the entropy per unit volume and
the chemical potential respectively, with $f=F/V$ the Helmholtz free energy per unit volume. 
The scalar curvature $R_o$ constructed out of $dl_o^2$ has the dimensions of volume and is therefore an appropriate 
candidate for the correlation volume. 

\section{The scalar curvature in closed and open thermodynamic systems}
\label{OC}

The scalar curvature constructed out of the metric of closed (Eq. (\ref{lineoriginal})) 
and open (Eq. (\ref{lineperuv}))
thermodynamic systems, when compared after transforming to the same coordinates, 
might generally not be the same. But for the cases of
our interest in this paper, namely when $c_v$ vanishes, the two are related, as we now show. 

To this end, we start by considering the Helmholtz free energy per unit volume of a fluid system, given as
\begin{equation}
f(T,\rho)=-\rho k_{B}T\log(e/\rho)+\rho h(T)-a\rho^{2}-\rho k_{B}T
\log(1-b\rho)+d_{1}\rho^{4}+d_{2}\rho^{6}~,
\label{freeperuv}
\end{equation}
where the first two terms on the right hand side of Eq. (\ref{freeperuv}) represent the ideal gas contribution
to the free energy, and $h(T)=-c_{v} k_{B} T \log(T/e)$. Here, $a$, $b$, $d_1$ and $d_2$ are constants. 
In $(T,v)$ coordinates, the Helmholtz free energy per unit volume takes the form
\begin{equation}
f(T,v)=-\frac{k_{B}T}{v} \log(ev)-
\frac{c_{v}k_{B}T}{v}\log(T/e)-\frac{a}{v^{2}}-\frac{ k_{B}T}{v}
\log\left(\frac{v-b}{v}\right)+\frac{d_{1}}{v^{4}}+\frac{d_{2}}{v^{6}}~,
\end{equation}
and results in the equation of state via $P = -\partial F/\partial V$, 
\begin{equation}
P = \frac{k_BT}{v-b} - \frac{a}{v^2} + \frac{3d_1}{v^4} + \frac{5d_2}{v^6}~.
\label{Pressgen}
\end{equation}
When $a=b=d_{1}=d_{2}=0$, Eq. (\ref{freeperuv}) describes the free energy per unit volume of the ideal gas,
while $d_{1}=d_{2}=0$ describes the Van der Waals (VdW) fluid. If $b=d_{2}=0$, the pressure
in Eq. (\ref{Pressgen}) has the same form as that of the RN-AdS black hole \cite{KubiznakMann} 
and $b=d_{1}=0$ represents the slowly rotating Kerr-AdS black hole in the same sense \cite{AKMS}, \cite{GKM}. 
While for finite non-zero values of $c_v$, these represent the RN-AdS fluid \cite{tapo3} or the
Kerr-AdS fluid, it will be important for us to consider the limit $c_{v}\rightarrow 0$, since the 
specific heat vanishes for both the RN-AdS black hole and the slowly rotating Kerr-AdS 
black hole. 

To compute the scalar curvature $R_c$, we write down the relevant metric
components from Eq. (\ref{lineoriginal}), 
\begin{equation}
g^{c}_{TT}(T,v)=\frac{C_{V}}{k_{B}T^{2}}, ~~~
g^{c}_{vv}(T,v)=-\frac{N}{k_{B}T}\left[\frac{\partial P(T,v)}{\partial
v}\right]_{T}=\frac{N}{k_{B}T}\left(\frac{k_{B} T}{(b-v)^2}+\frac{2 \left(15 d_{2}+6
d_{1} v^2-a v^4\right)}{v^7}\right)~.
\label{line1}
\end{equation}
The scalar curvature in terms of $T$ and $\rho$ reads
\begin{equation}
R_{c}(T,\rho) = \frac{2 k_{B} \rho   W(\rho)
\Big(k_{B} T+\rho W(\rho)\Big)}{C_{V} \Big(k_{B} T+2 \rho  W(\rho)\Big)^2}~,
\label{Rc}
\end{equation}
where $W(\rho)=(1-b \rho )^2 \left(6 d_{1} \rho ^2+15
d_{2} \rho ^4-a\right)$. 

Now for the open subsystem case, the metric components are calculated to be 
\begin{eqnarray}
g^{o}_{TT}(T,\rho)=\frac{\rho c_{v}}{T^{2}}~,~~
g^{o}_{\rho\rho}(T,\rho) = \frac{1}{k_{B} T}\left(\frac{k_{B}
T}{\rho  (1-b \rho )^2}+2  \left(-a+6 d_{1} \rho ^2+15 d_{2} \rho ^4\right)\right)~,
\label{line2}
\end{eqnarray}
and we call the scalar curvature of this open system as $R_o$. The expression for the normalized  scalar
curvature in this case is lengthy, but simplifies in the limit of
$c_{v}\rightarrow0$ to 
\begin{equation}
\lim_{c_{v}\rightarrow 0}\Big(R_{o}(T,\rho)c_{v}\Big)=\frac{2 W(\rho)
\Big(k_{B} T+\rho  W(\rho)\Big)}{\Big(k_{B} T+2 \rho  W(\rho)\Big)^2}~.
\label{Ro}
\end{equation}
Comparing Eqs. (\ref{Rc}) and (\ref{Ro}), we obtain
\begin{equation}
\frac{R_{c}C_{V}}{\rho }=\lim_{c_{v}\rightarrow 0}\left(R_{o}c_{v}k_{B}\right),
~~~\text{i.e.,} ~~~~R_{c}c_{v}=\lim_{c_{v}\rightarrow
0}\left(\frac{R_{o}c_{v}}{V}\right)~.
\label{cveq}
\end{equation}
Now, following our earlier
discussion, $R_{c}$ is dimensionless and $R_{o}$ carries the dimension of volume, so Eq. (\ref{cveq}) 
yields a relation consistent with dimensional analysis. 
In fact, we can substantiate this result further. As we have mentioned in the introduction, near criticality, 
$R_{o} \sim \xi^3$, with $\xi$ being the correlation length (the same is not true for $R_{c}$, which is
dimensionless). Then, we can write near criticality,
\begin{equation}
\lim_{c_v \to 0} \frac{\xi^3 c_v}{V} = K~,
\label{Keq}
\end{equation}
where $K=R_{c}c_v$ is a dimensionless quantity, independent of any scale of the system, and is in this sense universal. 
Therefore, a proper interpretation of $R_{c}c_v$ is that it is the (normalized) curvature of a corresponding open
thermodynamic system, per unit system volume. 

An important point to note here is that in $(T,V)$ coordinates, the metric obtained by
simply dividing that of Eq. (\ref{lineoriginal}) by the system volume, i.e.,
\begin{equation}
g_{TT} = \frac{C_V}{Vk_BT^2}~,~~g_{VV} = \frac{1}{Vk_BT}\left(\frac{\partial^2 F}{\partial V^2}\right)=
-\frac{1}{Vk_BT}\left(\frac{\partial P}{\partial V}\right)
\label{line2}
\end{equation}
gives the same scalar curvature $R_o$ for open subsystems (computed from Eq. (\ref{lineperuv})), 
as is readily checked. We note however that the form of the metric in Eq. (\ref{line2})
cannot be straightforwardly justified either from Eq. (\ref{lineoriginal}) or from Eq. (\ref{lineperuv}).
However, Eq. (\ref{line2}) will be important for two reasons. 

First, it is useful in situations where the exact analytical form of the 
Helmholtz free energy $F$ might be difficult to obtain, and one has to resort to approximations to 
write down the analytic form of the equation of state by which $P=P(V,T)$, 
within the limits of such approximations. Such a situation occurs in the study of 
Kerr-AdS black holes, which we will undertake in section \ref{KAdS}. 
Secondly, as is well known, the concept of a specific volume is somewhat challenged in 
black hole thermodynamics in extended phase space. Since the number of degrees of freedom $N$ of a 
black hole (which, in Planck units, is proportional to its horizon area) and its thermodynamic volume $V$
are both functions of the radius of the horizon, these are not independent. While for the
RN-AdS black hole, the specific volume appearing in its equation of state is $\rho = N/V$, the same is
not true for Kerr-AdS black holes, with the specific volume diverging for ultra-spinning cases \cite{AKMS}. 
Eq. (\ref{line2}) avoids this ambiguity and provides a generic way to compute the scalar curvature
for black holes, with the appropriate dimension of volume. Henceforth, Eq. (\ref{line2}) will be our line element 
for the black hole cases, and by a slight abuse of notation, we will continue to refer to the 
scalar curvature obtained from the metric of Eq. (\ref{line2}) as $R_o$, for the black hole cases, 
so as not to clutter the notation.

We pause here to mention a few important details. First of all, we note that close to the second order 
critical point, fluid systems are solely characterized by the critical exponents (mean field exponents 
in classical systems that we consider), and therefore exhibit certain universal properties. For example, 
for the VdW fluids with finite $c_v$, the quantity $R_{o}c_v/V \sim t^{-2}$ is universal, 
near criticality, with $t = T/T_c -1$, $T_c$ being the critical temperature \cite{Rupp}. Eq.(\ref{cveq}) 
is on the other hand a relation that is valid for systems with vanishing $c_v$, even away from criticality. 
Also, we note that for two dimensional parameter spaces,
$R_{o}c_v$ is the {\it only} normalized scalar that has dimensions of volume, and scales at the correlation length
near criticality. Hence, it is natural to demand that this normalized scalar curvature, which carries 
information about the scale of the system, carries the limits of applicability too, like theories with non-zero
heat capacity that we mentioned in the introduction. For example, whenever $R_{o}c_v$ becomes comparable to,
or smaller than the specific volume, 
the formalism itself should be understood to have broken down. There are some further compatibility features, 
as we will discuss in the next section. 

For the moment, as an example, we consider the simplest non-trivial classical fluid system that is amenable
to a geometric analysis, namely the VdW fluid. The VdW equation of state and its reduced form
read
\begin{equation}
P = \frac{k_BT}{\left(v-b\right)} - \frac{a}{v^2}~,~~
P_r = \frac{8T_r}{\left(3v_r - 1\right)} - \frac{3}{v_r^2}~,
\label{VdW}
\end{equation}
respectively, where $v$ is the volume per molecule, and the reduced pressure, volume and temperature are 
given as $P_r = P/P_c$, $v_r = v/v_c$ and $T_r=T/T_c$, with $P_c = a/(27b^2)$, $v_c = 3b$ and 
$T_c = 8a/(27bk_B)$ denoting the critical values. The scalar curvature computed from Eq. (\ref{lineoriginal}) 
is called $R_{c}$, and given by \cite{WeiLiuMann}
\begin{equation}
R_{c} = \frac{k_B\left(3v_r -1\right)^2\left(\left(1-3v_r\right)^2 - 8T_rv_r^3\right)}{2C_V\left(\left(1-3v_r\right)^2 - 4T_rv_r^3\right)^2}~.
\end{equation}
On the other hand, the scalar curvature computed from Eq. (\ref{lineperuv}) (or equivalently from Eq. (\ref{line2})) is 
$R_{o} = {\mathcal A.B}$, where we have defined \cite{RSSS}
\begin{eqnarray}
{\mathcal A} &=& \frac{3bv_r\left(3v_r-1\right)}{18c_v\left(\left(1-3v_r\right)^2 - 4T_rv_r^3\right)^2}~,
\nonumber\\
{\mathcal B} &=& 9\left(3v_r-1\right)\left(8T_rv_r^3-\left(1-3v_r\right)^2\right)+4c_vT_rv_r
\left(1-9v_r + 27v_r^2 - 27v_r^3 + 8T_rv_r^3\right)~.
\end{eqnarray}
Clearly then we have the relation $\lim_{c_v \to 0}R_{o}c_v/V = R_{c}c_v~$, confirming Eq. (\ref{cveq}). 
While this holds away from criticality, 
note that the coefficient of $c_v$ in the second term of ${\mathcal B}$ vanishes close to
criticality, i.e in the limit $T_r,v_r \to 1$. This is as we have mentioned before -- close to the second order
critical point, $R_{o}c_v/V$ is universal, even for non-zero heat capacity. 

We have thus established that for an open thermodynamic fluid system, the scalar curvature on the parameter space is
related to that of the same system considered as closed, by a factor of the system volume, in the limit that the
specific heat goes to zero. As mentioned before, \cite{RuppOriginal} used Eq. (\ref{lineoriginal}) 
as the line element, and inserted the system volume appropriately, in order to obtain a
scalar curvature of appropriate volume dimension. What we have shown here is that this will yield the scalar
curvature of the corresponding open thermodynamic system, in the limit of vanishing $c_v$. 
In this context, we mention that similar results hold for the
RN-AdS fluid \cite{tapo3} which is a two-parameter family of fluid solutions having the same equation of
state as the RN-AdS black hole. Of course for real fluids (VdW or otherwise), this is more of a mathematical curiosity,
as $c_v$ for such fluids is always non-zero. 

\section{RN-AdS black holes}
\label{RNAdS}

Recall that in \cite{Ray}, it was proposed that a varying cosmological constant can be identified with the
pressure of a black hole, with the volume of the event horizon being its conjugate. In this
formalism, the mass of the black hole is identified with its enthalpy (rather than its internal
energy), and this can be shown to satisfy a consistent Smarr relation. 
Following \cite{Ray}, the work of \cite{KubiznakMann} derived a relation between the
pressure, volume and temperature of an RN-AdS black hole with electric charge $Q$. 
This shows similar phase behaviour as that of the VdW fluid and is often used fruitfully to 
understand the coarse grained structure of these black holes. 

As we have already mentioned, a verbatim application of thermodynamic geometry in black hole 
thermodynamics is rendered difficult by its very nature. Recall that in the static case considered here, the number
of degrees of freedom $N$ and the system volume $V$ are not independent, and both depend on the 
radius of the event horizon \cite{AKMS}. One can nonetheless construct $R$ with appropriate dimensions
of volume, starting from the Helmholtz free energy $F$ per unit thermodynamic volume $V$, 
as is the case with ordinary thermodynamic systems. 
With $r_+$ denoting the radius of the event horizon, and the black hole volume $V=(4/3)\pi r_+^3$, this reads 
from Eq. (3.29) of \cite{KubiznakMann},
\begin{equation}
\frac{l_p^2}{\hbar c}\frac{F}{V} = \left(\frac{\pi}{6}\right)^{1/3}\frac{Q^2}{V^{4/3}} +\left(\frac{3}{32\pi}\right)^{1/3}\frac{1}{V^{2/3}}
- \left(\frac{9\pi}{16}\right)^{1/3}\frac{T}{V^{1/3}}~,
\label{Fperuv}
\end{equation}
where  $l_p$ is the Planck length, $\hbar$ is Planck's constant and $c$ the speed of light. Here, $Q$
and $T$ have the dimensions of length and inverse length respectively, and 
the physical charge and temperature are given in terms of $Q$ and $T$ as $Q^2G/(4\pi\epsilon_0 c^4)$ and 
$\hbar cT/k_B$, respectively, with $G$ being the four-dimensional gravitational constant and $\epsilon_0$
the permittivity of vacuum. 
Here, $v=2l_p^2r_+$ is the specific volume with its critical value
$v_c = 2l_p^2\sqrt{6}Q$. In terms of the reduced variable $v_r = v/v_c=V_r^{1/3}$, 
with $V_r = V/V_c$ (the critical values of the temperature, pressure
and volume being $T_c = \sqrt{6}/(18\pi Q),~P_c = 1/(96\pi Q^2),~V_c = 8\sqrt{6}\pi Q^3$), the equation of
state is 
\begin{equation}
P_r = \frac{8T_r}{3v_r} - \frac{2}{v_r^2} + \frac{1}{3v_r^4}
\label{eos}
\end{equation}
If we use the Helmholtz free energy $F$ instead of the free energy per 
unit volume of Eq. (\ref{Fperuv}), we obtain \cite{WeiLiuMannPRL} using Eq. (\ref{lineoriginal}),
\begin{equation}
R_{c}c_v = \frac{\left(3v_r^2-1\right)\left(3v_r^2-4T_rv_r^3 -1\right)}{2\left(1-3v_r^2 + 2T_rv_r^3\right)^2}~.
\label{scalarWLM}
\end{equation}
On the other hand, using Eq. (\ref{line2})\footnote{Note that if we used Eq. (\ref{Fperuv}) in 
Eq. (\ref{lineperuv}) with $\rho = 1/v$, the scalar curvature evaluates to $\sqrt{6}Ql_p^2$, 
which does not capture any critical behavior, and the analogy to fluid systems is lost. We therefore use
Eq. (\ref{line2}) for computing the scalar curvature.}, the scalar curvature per unit specific volume is given in
the limit $c_v \to 0$ by 
\begin{equation}
\frac{R_{o}c_v}{v}=\left(\frac{Q^2}{l_p^2}\right)
\frac{2\pi v_r^2\left(3v_r^2-1\right)\left(3v_r^2-4T_rv_r^3 -1\right)}{\left(1-3v_r^2 + 2T_rv_r^3\right)^2}~.
\label{scalar}
\end{equation}
Clearly then, as mentioned in Eq. (\ref{cveq}), $R_{c}c_v=\lim_{c_v\to 0} R_{o}c_v/V$. 
This is in lines with our previous argument : if one uses the line element of Eq. (\ref{lineoriginal}), then 
the scalar curvature computed from this agrees with the one obtained using Eq. (\ref{line2}) by
an overall factor of the system volume, in the limit $c_v \to 0$, which is indeed the case here. 
Note that the scalar curvature $R_{o}c_v$ (with appropriate dimensions of volume) is {\it not}
universal, but the curvature per system volume is. 

With a dimensionally consistent definition (Eq. (\ref{scalar})), we are in a position to examine the validity of
the geometric approach to black hole thermodynamics in this case. There are a few important issues that
need to be examined. First of all, we emphasize that in a coarse-grained approach, any quantity that has
magnitude lower than the specific volume loses significance, as it becomes lower than the lowest scale
allowed by the system. Indeed, as discussed in details in \cite{RSSS}, for the VdW fluid system,
this restricts the usage of the geometric methods discussed there (in a somewhat different context of locating first 
order phase transitions and the Widom line \cite{Widom} in the supercritical region 
using geometric methods) to some suitable limits near criticality. 
Namely, in the sub-critical region, one is restricted to $T_r \gtrsim 0.8$ and in the super-critical region, 
to $P_r \lesssim 10$. It is important to understand such restrictions in the current scenario,
which would imply that the right hand side of Eq. (\ref{scalar}) should greater than unity, namely that we
should demand $R_{o}c_v/v > 1$. 

A further significant issue is that from the equation of state of Eq. (\ref{eos}), one readily finds that in
order to avoid negative temperatures, there exists a minimum value of the reduced volume for a 
given pressure, namely 
\begin{equation}
v_r^{min} = \frac{1}{\sqrt{3 + \sqrt{9 + 3P_r}}}~,~{\rm i.e}~,~~v^{min}= \frac{v_c}{\sqrt{3 + \sqrt{9 + 3P_r}}}~.
\label{vrmin}
\end{equation}
This would imply the possibly weaker constraint $R_{o}c_v/v^{min} >1$. 

Finally, we note that the geometric formalism is based on a Gaussian approximation. This implies that
the entropy of the system is expanded about an equilibrium (extremal) value, and one retains only the 
lowest order correction to the same. As explained in \cite{Rupp} (see also the pegagogical exposition
in \cite{RuppAJP}), this means that one should 
reasonably expect that the scalar curvature is much less than the system volume, which in this
case translates to $R_{o}c_v \ll V$. Near a second order phase transition, this condition is violated, and 
signals the expected breakdown of the Gaussian approximation near criticalitly. 
This constraint is often overlooked in standard analyses, namely because
the system volume can always be chosen to be large. However the subtlety for RN-AdS black holes
is that this is also {\it determined} by the radius of the outer event horizon, which also fixes the specific volume. 
Specifically then, from our previous discussion, we expect that the dimensionless quantity on 
the right hand side of Eq. (\ref{scalarWLM}) is much less than unity. 
To summarize, the conditions that we expect to hold in our coarse grained analysis are 
\begin{equation}
\frac{R_{o}c_v}{v} > 1~,~~\frac{R_{o}c_v}{v^{min}}>1~,~~\frac{R_{o}c_v}{V} \ll 1~,
\label{conditionsmain}
\end{equation}
where the second condition is a weaker one that follows naturally from the first, but will be useful
for us in what follows. 

We will now examine these constraints one by one. In this section, we will study Eq. (\ref{conditionsmain})
for RN-AdS black holes. In the next section, we will comment upon these in the context of Kerr-AdS black holes. 
To make contact with existing literature, we recall that in \cite{SS} it was shown that for the RN-AdS black
hole, one can derive an analytical formula for the coexistence curve. In terms of the reduced temperature,
the reduced volume on the liquid ($v_r^l$) and gas ($v_r^g$) sides at the first order phase transition where the Helmholtz
free energy becomes equal for the two phases can be written as \cite{SS}
\begin{equation}
v_r^{l,g} = \frac{\sqrt{1+y} \mp \sqrt{3y}}{1-2y}~,~~y = \cos\left(\frac{1}{3}\left(\arccos(1-T_r^2) + \pi\right)\right)~.
\label{saturationv}
\end{equation}
It is then readily checked that in the limit $T_r \to 0$, $v_r^{min} = v_r^l|_{T_r\to 0} = 1/\sqrt{6}$.
At this point, we recall that for the VdW fluid, the reduced minimum volume is numerically equal to $1/3$, independent
of the pressure, which is the hard sphere cutoff in that case. In our case, 
to make physical sense of the scalar curvature, one is thus forced to restrict to situations where $|R_{o}|c_v >v^{min}$.
\begin{figure}[h!]
\begin{minipage}[b]{0.5\linewidth}
\centering
\includegraphics[width=2.5in,height=1.6in]{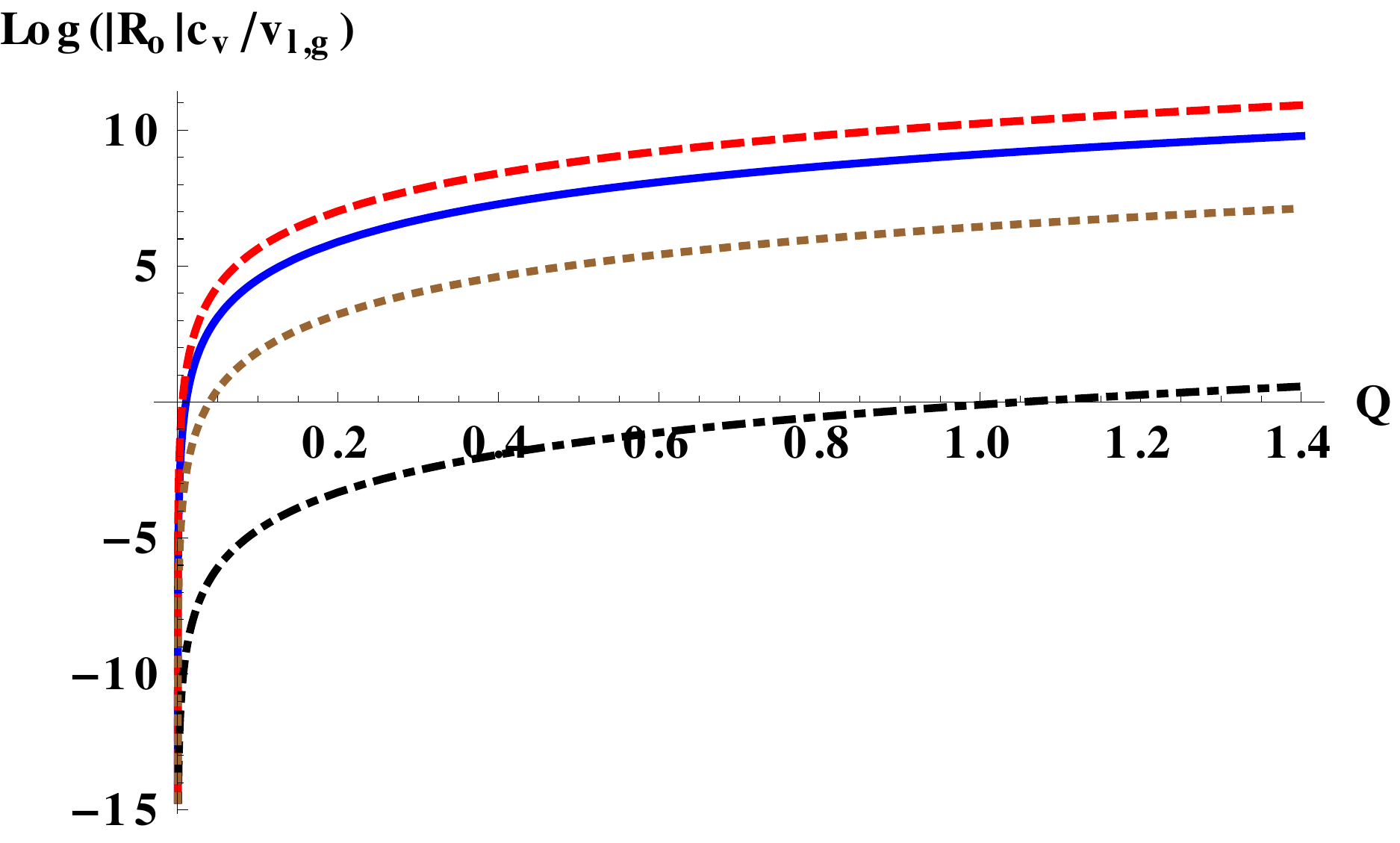}
\caption{Saturation values of $\log(|R_{o}|c_v/v)$ as a function of $Q$ for different $T_r$. The solid blue and the dashed 
red lines correspond to $\log(|R_{o}|c_v/v^l)$ and $\log(|R_{o}|c_v/v^g)$ respectively, for $T_r=0.99$. 
The dot-dashed black and the dotted brown lines correspond to the same quanties for $T_r=0.7$.}
\label{fig1}
\end{minipage}
\hspace{0.4cm}
\begin{minipage}[b]{0.5\linewidth}
\includegraphics[width=2.5in,height=1.6in]{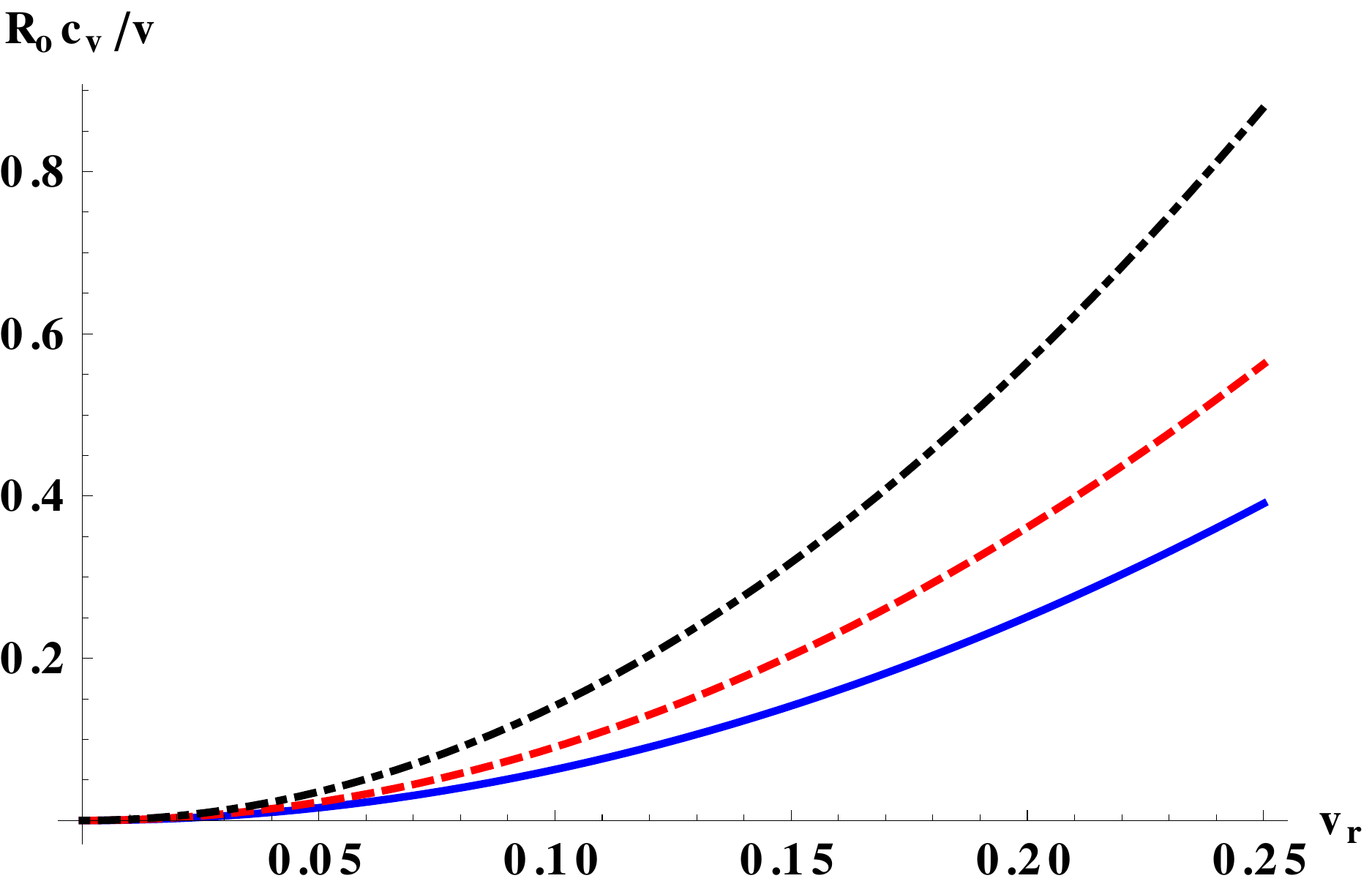}
\caption{Supercritical values of $R_{o}c_v/v$ as a function of $v_r$ for $T_r=2$ for different charges. 
The solid blue, dashed red and dot-dashed black curves correspond to $Q = 1$, $1.2$ and $1.5$ respectively}
\label{fig2}
\end{minipage}
\end{figure}
To make the above more concrete, we first write Eq. (\ref{scalar}) in terms of $P_r$ and $v_r$, yielding 
\begin{equation}
R_{o}c_v = \frac{32\sqrt{2}}{3\sqrt{3}}\frac{\pi Q^3 v_r^3\left(1-3v_r^2\right)\left(1+3P_rv_r^4\right)}{\left(1-2v_r^2 + P_rv_r^4\right)^2}~.
\end{equation}
Corresponding to the minimum value of $v$, i.e., $v^{min} = v_cv_r^{min}$, we obtain
\begin{equation}
\frac{R_{o}c_v}{v^{min}}= \frac{2\pi Q^2\left(\sqrt{9+3P_r} - 3\right)}{3l_p^2 P_r}
\end{equation}
The condition $R_{o}c_v > v^{min}$ then implies that 
\begin{equation}
P_r < \frac{4\pi Q^2}{3l_p^2}\left(\frac{\pi Q^2}{l_p^2} - 3\right)~,~~{\rm and}~~\frac{Q}{l_p} > \frac{\sqrt{3}}{\sqrt{\pi}}~.
\label{conQPr}
\end{equation}
Note that the constraint of Eq. (\ref{conQPr}) is minimal, i.e., based on the minimum volume at a given
pressure. It is useful to check the validity of the geometric approach in more general cases, for example
at the saturation curve and in the super-critical region. For this, one has to check that the quantity $|R_{o}|c_v/v > 1$ 
in both these regions, with the constraint on the charge given in Eq. (\ref{conQPr}). 

In Fig. (\ref{fig1}), we plot the quantity $\log(|R_{o}|c_v/v)$ as a function of the charge $Q$ (after setting $l_p=1$), 
for different values of the reduced temperature, at saturation, using Eq. (\ref{saturationv}). In the figure, the solid blue and the dashed 
red lines correspond to $\log(|R_{o}|c_v/v^l)$ and $\log(|R_{o}|c_v/v^g)$ respectively, for $T_r=0.99$. 
The dot-dashed black and the dotted brown lines correspond to $\log(|R_{o}|c_v/v^l)$ and $\log(|R_{o}|c_v/v^g)$ respectively, 
for $T_r=0.7$. We note that for the last case, the geometric analysis becomes invalid on the liquid
side of the saturation curve, even for $Q\sim 1$, which satisfies the constraint of Eq. (\ref{conQPr}). 
Broadly, our analysis implies that at saturation, we are restricted to $T_r \gtrsim 0.75$ for the geometric analysis
to be valid. 

In Fig. (\ref{fig2}), we plot the quantity $R_{o}c_v/v$ in the super-critical region $T_r=2$ as a function of $v_r$, for different 
values of the charge, again setting $l_p=1$. The solid blue, dashed red and dot-dashed black curves correspond to 
$Q = 1$, $1.2$ and $1.5$ respectively, all of which satisfy the bound given in Eq. (\ref{conQPr}). 
We see however that in the region $v_r<0.25$, all the curves indicate $Rc_v/v <1$, i.e the
geometric analysis is strictly invalid in these regions. 
\begin{figure}[h!]
\begin{minipage}[b]{0.5\linewidth}
\centering
\includegraphics[width=2.5in,height=1.6in]{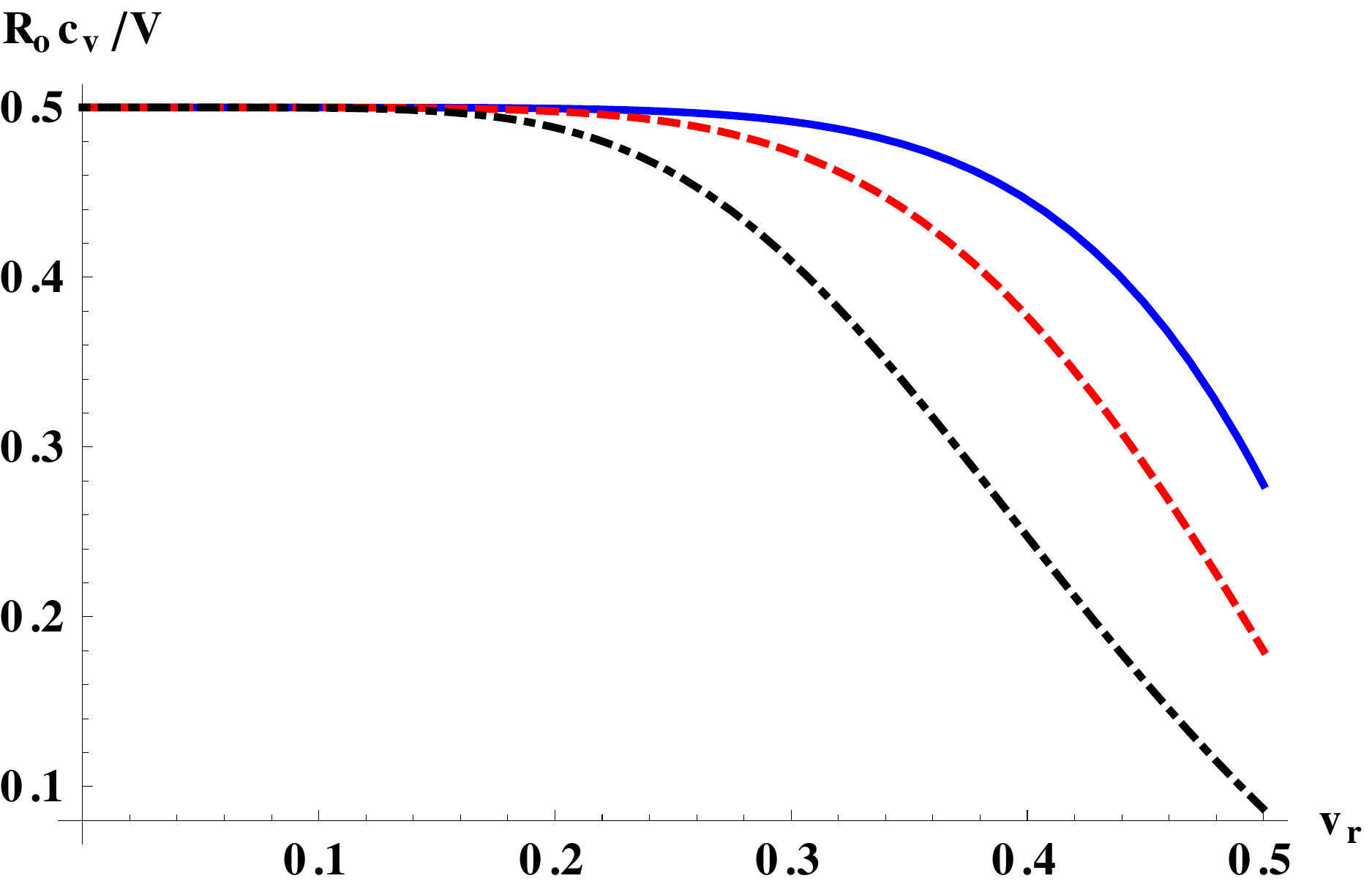}
\caption{Supercritical values of $R_{o}c_v/V$ as a function of $v_r$ for different $T_r$.
The solid blue, dashed red and dot-dashed black correspond to $T_r=2$, $4$ and $10$, respectively.}
\label{fig3}
\end{minipage}
\hspace{0.4cm}
\begin{minipage}[b]{0.5\linewidth}
\includegraphics[width=2.5in,height=1.6in]{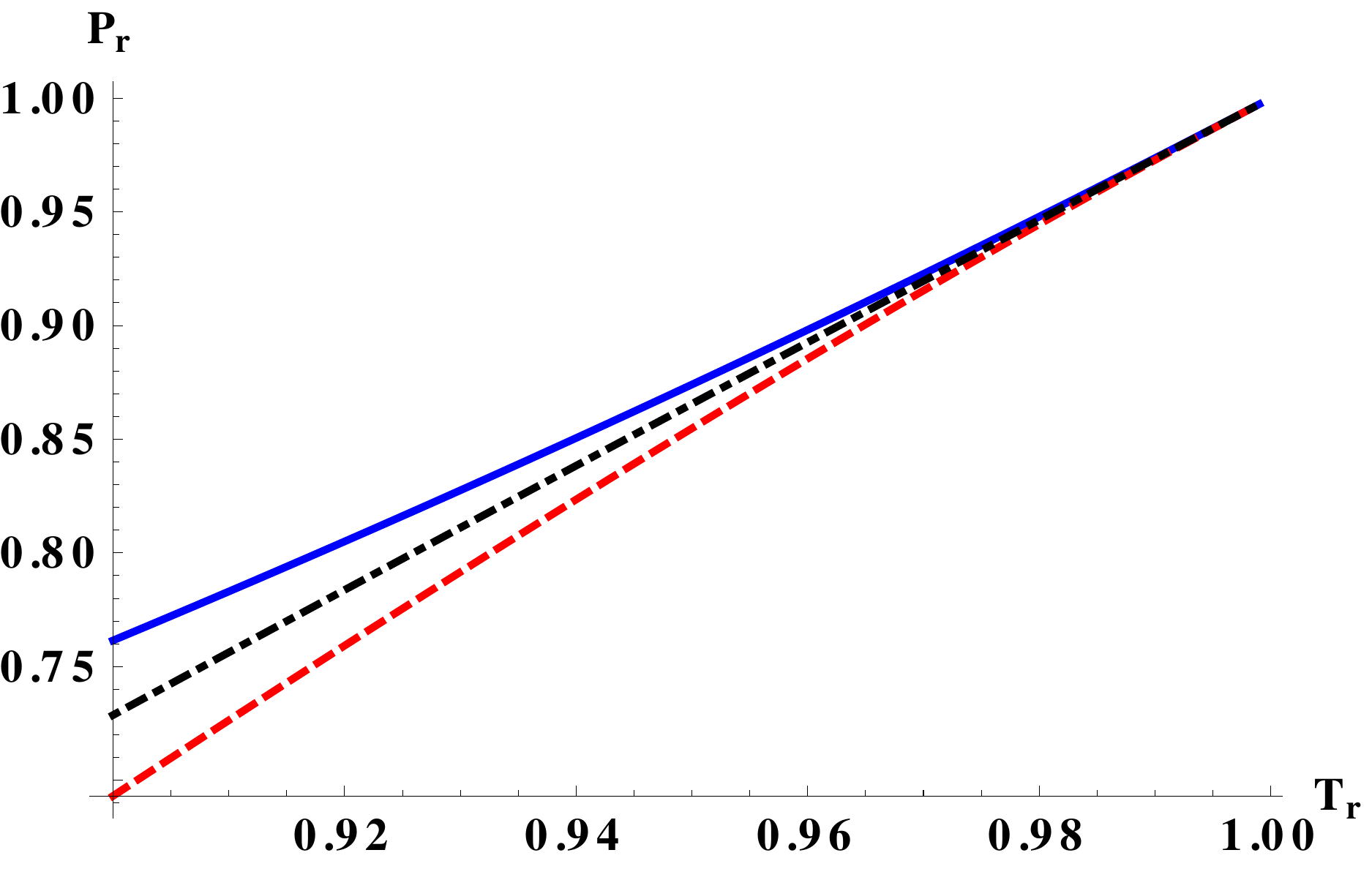}
\caption{Curves depicting equality of various quantities on the $T_r - P_r$ plane. The solid blue
line corresponds to the phase coexistence curve that is arrived at by the Maxwell construction
in the $P - V$ plane. The dashed red curve depicts the locus of equality of the quantity $R_{o}c_v$ of 
and the dot-dashed black curve to the equality of $R_{c}c_v$.}
\label{fig4}
\end{minipage}
\end{figure}
In Fig. (\ref{fig3}), we plot $R_{o}c_v/V$ 
as a function of $v_r$ for different values of $T_r$. The solid blue, dashed red and dot-dashed black here
correspond to $T_r=2$, $4$ and $10$, respectively. We note that for all these values of $T_r$, $R_{o}c_v \sim 0.5$,
for small $v_r$, a fact that readily follows by taking the limit of small $v_r$ in Eq. (\ref{scalarWLM}). 
This seems to indicate from our arguments above that the geometric formalism possibly does not have 
a proper physical interpretation in these regions. In fact, it can be checked that if we strictly enforce $R_{o}c_v \ll V$, 
the allowed region of parameter space is indeed very small. 

Finally, it is of interest to consider how the correlation length indicates first order phase transitions near criticality. 
Following \cite{Widom}, it was suggested in \cite{RSSS} that close to the second order critical point, first order
phase transitions can be predicted from the equality of the correlation length in the liquid and the gas phases. 
Since the scalar curvature is proportional to the correlation volume, the equality of this in the liquid and gas
phases was conjectured to predict first order phase transitions, close to criticality. 
In \cite{RSSS}, this conjecture was checked for simple fluids (say ideal gases) that closely follow the VdW equation of state. 
The computation becomes somewhat complicated in these cases, due to the fact that $c_v$ is strictly
temperature dependent and does not equal the ideal gas value on the liquid side. RN-AdS black holes provide
a good testing ground for this conjecture bypassing this difficulty, since $c_v$ vanishes identically. 
Importantly, in this exercise since we compare two expressions for $R_{o}c_v$, the charge appearing in 
this (as follows from Eq. (\ref{scalar})) cancels out. 
In Fig. (\ref{fig4}), we have constructed the phase coexistence curve on the $T_r - P_r$ plane. The solid blue
line corresponds to the phase coexistence curve of \cite{SS} that is arrived at by the Maxwell construction
in the $P - V$ plane. The dashed red curve depicts the locus of equality of the quantity $R_{o}c_v$ of 
Eq. (\ref{scalar}) and the dot-dashed black curve to the equality of $R_{c}c_v$ of Eq. (\ref{scalarWLM}). Clearly,
all the curves merge close to criticality, verifying the conjecture of \cite{RSSS} for the RN-AdS black 
hole.\footnote{Curiously, we find that equality of $R_{c}c_v$ provides a somewhat better approximation to the 
Maxwell construction away from criticality.} 

\section{Kerr-AdS black holes}
\label{KAdS}

We will now briefly consider the extended phase space geometry of the four dimensional Kerr-AdS black holes, 
whose thermodynamics was considered in \cite{GKM},\cite{AKMS}. Exact computations in this case are known 
to be complicated, and we will simply comment upon an approximate equation of state obtained in the slow rotation 
limit of the black hole. This is given in (Eq. (3.35) of \cite{AKMS})
\begin{equation}
P = \frac{T}{v} - \frac{1}{2\pi v^2} + \frac{48J^2}{\pi v^6} - \frac{384J^4\left(8\pi Tv + 7\right)}{\pi v^{10}\left(\pi Tv + 1\right)^2}
+ \frac{36864J^6\left(13\pi Tv + 11\right)}{\pi v^{14}\left(\pi TV + 1\right)^3}~+ {\mathcal O}(J^8)~,
\label{KAdSeos}
\end{equation}
where $J$ is the angular momentum of the black hole and $P$, $T$ and $v$ denote the pressure, temperature and
specific volume respectively, as before. Note that similar to the RN-AdS case, we have $V = (1/6)\pi v^3$. 

We will mainly focus on the equation of state up to ${\mathcal O}(J^2)$, i.e., retaining the first three terms on the
right hand side of Eq. (\ref{KAdSeos}), 
and note that the specific heat at constant volume, $c_v=0$ to this order \cite{AKMS},\cite{GKM}.
In a similar fashion as illustrated in \cite{tapo3}, the geometry of the system at this order is conveniently studied by assuming a 
Kerr-AdS fluid that has an equation of state of the form 
\begin{equation}
P =  \frac{T}{v} - \frac{a}{v^2} + \frac{d}{v^6}~,
\label{KAdSfluid}
\end{equation}
where $a$ and $d$ are two constants of appropriate dimensions, that characterize the departure 
of the fluid from an ideal gas form. For the slowly rotating KAdS black hole,
we have $a=1/(2\pi)$ and $d= 48J^2/\pi$. 
Following our discussion in section \ref{OC}, one can check that
the free energy per unit volume of the KAdS fluid can be written as
\begin{equation}
f(T,\rho) = f_{id} - a\rho^2 + \frac{d}{5}\rho^6~,
\label{freeenergy}
\end{equation}
where $\rho =1/v$, and $f_{id}$ is the ideal gas part of the free energy, given by the first two terms on
the right hand side of Eq. (\ref{freeperuv}). 
The critical values of the thermodynamic quantities are
\begin{equation}
T_c = \frac{8}{5k_B}\left(\frac{a^5}{15d}\right)^{1/4}~,~v_c = \left(\frac{15d}{a}\right)^{1/4}~,~
P_c = \frac{2 a^{3/2}}{3 \sqrt{15d}}
\label{KAdScrit}
\end{equation}
Denoting $T = T_rT_c$, $v = v_rv_c$, $P=P_rP_c$, the reduced form of the equation of state is 
\begin{equation}
P_r = \frac{12T_r}{5v_r} - \frac{3}{2v_r^2} + \frac{1}{10v_r^6}~.
\label{reducedKAdS}
\end{equation}
$R_o$ is computed using Eq. (\ref{line2}), and we find that 
\begin{equation}
\lim_{c_v\to0}\frac{R_{o}c_v}{V} = \frac{\left(1-5v_r^4\right)\left(1-5v_r^4+8T_rv_r^5\right)}
{2\left(1-5v_r^4+ 4T_rv_r^5\right)^2}~=R_c c_v~,
\label{RopenKAdS}
\end{equation}
with $R_c$ computed from Eq. (\ref{lineoriginal}). 
Note that the minimum volume for a given pressure here has a more complicated expression
than that in the RN-AdS case, but in the limit $P_r\to 0$, this equals $1/(15)^{1/4}$ and goes to zero 
as $P_r \to \infty$. This in particular implies that much like the RN-AdS case, there is a lower bound
on $J$ below which the geometric analysis will break down. 

The analysis above is based on a slow rotation approximation where one only keeps
terms up to ${\mathcal O}(J^2)$ in the expression for the pressure in Eq. (\ref{KAdSeos}). 
Indeed, the general case, without such an approximation, is difficult to handle
analytically, the broad reason being that the exact dependence of the outer horizon radius as a function
of the angular momentum is difficult to obtain. Hence, the free energy has to be computed order by order 
via a perturbation expansion of the outer horizon radius in terms of the rotation parameter (for an elaboration of this method,
see \cite{AKMS}). Fortunately however, in \cite{WCL}, an exact expression for the critical point
was given via the temperature - entropy criticality and it was found that the 
critical point obtained in the  ${\mathcal O}(J^2)$ expansion of the equation of state in 
Eq. (\ref{KAdSeos}) was a reasonably good approximation, with the absolute values of the relative deviations
in the critical thermodynamic quantities being maximally $\sim 2.2\%$. This can be 
gleaned from the approximate values of the critical thermodynamic variables obtained from Eq. (\ref{KAdSeos}) 
at ${\mathcal O}(J^2)$, and their exact values \cite{WCL}, 
which read (with a superscript $e$ denoting the exact values),
\begin{equation}
\left(T_c, v_c, P_c\right) = \left(\frac{2^{3/4}}{5^{5/4} \sqrt{3} \pi  \sqrt{J}}, 10^{1/4} 2 \sqrt{3} \sqrt{J},
\frac{1}{36 \sqrt{10} \pi  J}\right)~,~~
\left(T^e_c, v^e_c, P^e_c\right) = \left(\frac{0.04175}{\sqrt{J}}, 6.04736\sqrt{J},\frac{0.00286}{J}\right)~.
\label{approxexact}
\end{equation}

Note that at ${\mathcal O}(J^2)$, one can begin with the {\it correct} constants, i.e., adjust $a$ and $d$ in Eq. (\ref{KAdSfluid}) 
so that two of the critical values above are exact. Namely, we use the expressions for the critical values of
$T_c$ and $v_c$ from Eq. (\ref{KAdScrit}), and equate them to $T^e_c$ and $v^e_c$ given 
in the second relation of Eq. (\ref{approxexact}). 
Then, instead of $a = 1/(2\pi)=0.159$ and $d = 48J^2/\pi=15.279J^2$, we obtain the slightly different values 
$a^e = 0.1578$ and $d^e = 14.0694J^2$. 
With our chosen values of $a=a^e$ and $d=d^e$, the reduced form of the equation of state in Eq. (\ref{reducedKAdS})
is of course unchanged. Then from Eq. (\ref{KAdScrit}),
we obtain the critical pressure $P_c = 0.002877/J$, with the absolute value of the relative error
$(P^e_c - P_c)/P^e_c\sim 0.66\%$, indeed a small deviation from the exact value. It is thus tempting 
to think that near criticality, the effect of the exact equation of state of the KAdS black hole is to effectively ``renormalize'' the 
constants $a$ and $d$ appearing in Eq. (\ref{KAdSfluid}), but we do not have a convincing proof for this. 
\begin{figure}[h!]
\centering
\includegraphics[width=2.8in,height=2.0in]{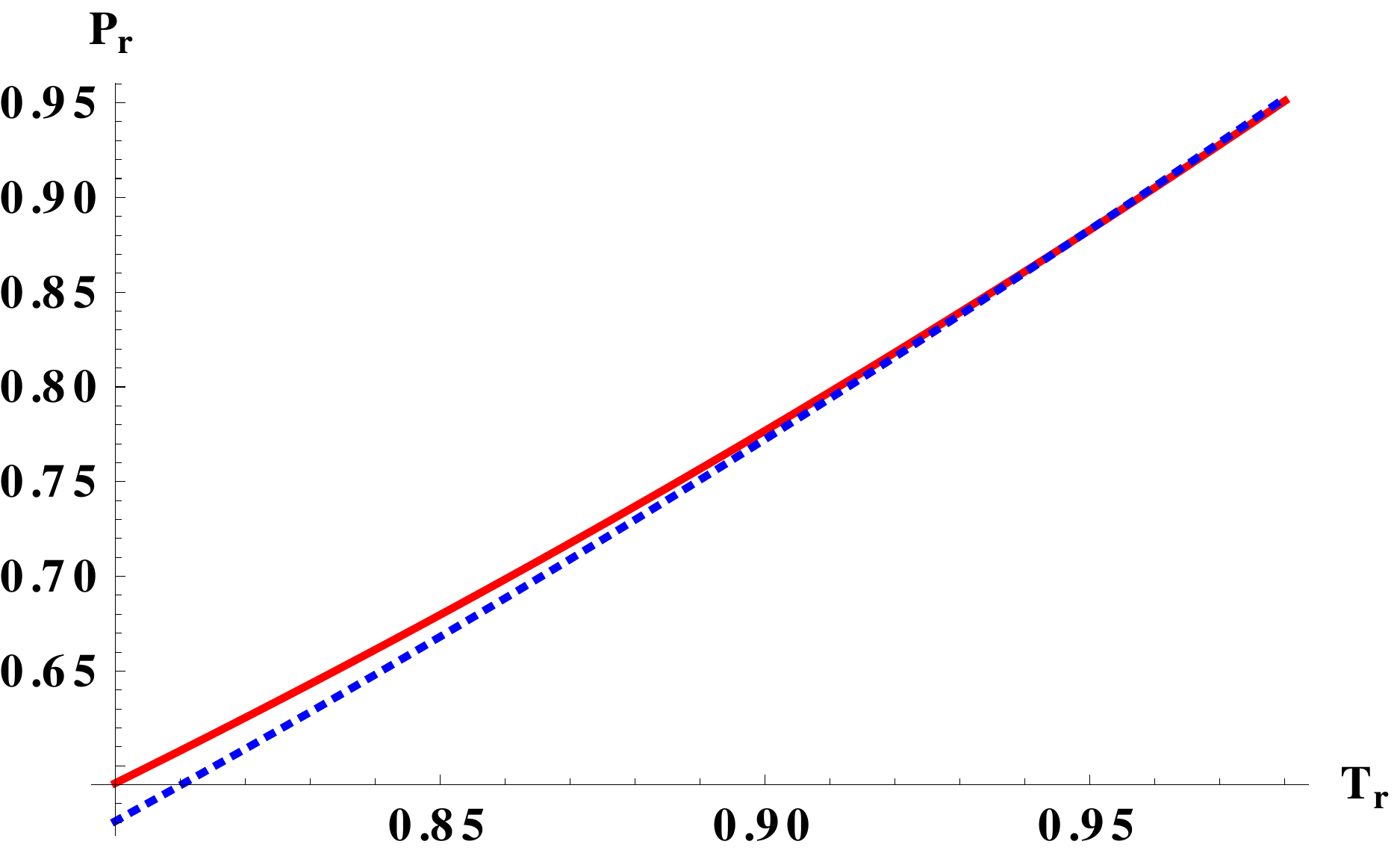}
\caption{Coexistence curve for the Kerr-AdS fluid on the $T_r-P_r$ plane. The solid red line is the fit 
given in \cite{WCL} and the dotted blue line is the approximate curve obtained from Eq. (\ref{KAdSfluid}).}
\label{fig5}
\end{figure}

Now, instead of the above, if we use the equation of state in Eq. (\ref{KAdSeos}) up to ${\mathcal O}(J^6)$, and 
make a reasonable assumption that the critical values of the thermodynamic quantities are 
now given by their exact values appearing in the second relation of Eq. (\ref{approxexact}), 
then we obtain the reduced equation of state at this order, 
\begin{equation}
P_r = \frac{2.41T_r}{v_r} -\frac{1.52}{v_r^2} + \frac{3.43\times 10^{-1}}{v_r^6} - \frac{6.49\times 10^{-4}\left(7+6.34 T_r v_r\right)}
{v_r^{10}\left(1 + 0.79 T_rv_r\right)^2} + \frac{4.68\times 10^{-5}\left(11+10.31T_rv_r\right)}{v_r^{14}\left(1 + 0.79 T_rv_r\right)^3}~.
\label{fulleq}
\end{equation}
Clearly then, close to criticality where $v_r$ and $T_r$ are ${\mathcal O}(1)$, terms higher than ${\mathcal O}(J^2)$
in the equation of state of Eq. (\ref{KAdSeos}) are strongly suppressed, and hence the  quadratic order equation 
of state of Eq. (\ref{KAdSfluid}) is sufficiently robust. This can be further substantiated as follows. 
Using Eq. (\ref{reducedKAdS}), one can numerically
plot the coexistence curve in the $T_r-P_r$ plane. In Fig. (\ref{fig5}), we show this with the dotted blue line
corresponding to the coexistence curve obtained via the Maxwell construction. 
In this figure, the analytical coexistence curve given in Eq. (43) of \cite{WCL}
is shown by the solid red line. One can clearly see that close to criticality, i.e., for $T_r \gtrsim 0.85$, 
the two curves almost merge, implying that in this region, the small $J$ equation of state provides an excellent 
approximation to the exact one. 

In fact, our analysis for small $J$ indicates that the geometric method is valid only in the region close to criticality, for
the Kerr-AdS fluid case. We arrive at this conclusion by computing $R_o c_v/v$ using Eq. (\ref{RopenKAdS}). At
saturation, we find that in order for the first condition of Eq. (\ref{conditionsmain}) to be true on the
liquid side of the saturation curve, we require, for $J=0.1$, $T_r \gtrsim 0.91$. For smaller values of $J$,
we are restricted even closer to the critical point. Since the small $J$ approximation is excellent near
criticality as we have just discussed, we can conclude that our results are robust and will not change
significantly if one relaxes the assumption of small $J$. Although this analysis is justified for
the Kerr-AdS fluid, we note that for the Kerr-AdS black hole, it is somewhat challenged, 
as the notion of the specific volume being the volume per degree 
of freedom does not hold \cite{AKMS}. However, purely from the equation of state 
for the slowly rotating Kerr-AdS black hole of Eq. (\ref{KAdSeos}), if we identify $v$ as an analog
of the specific volume, then the interpretation of  $R_o c_v/v$ will be similar to that of the Kerr-AdS fluid. 

\section{Summary and Discussions}
\label{Conclusions}

In the study of classical phase transitions driven by thermal fluctuations, the formalism of thermodynamic geometry 
has been popular of late. This Riemannian geometric formalism begins by constructing a metric on the parameter manifold
of the theory, and ultimately relates the Ricci scalar curvature of this manifold to the correlation length, near
criticality. For two-dimensional parameter manifolds where the Riemann tensor has a single independent component,
the Ricci scalar fully characterizes the curvature of the manifold and in that sense uniquely specifies the 
geometry. The conventional approach then is to treat the system as an open system, whose volume is held fixed,
and this is in turn in equilibrium with a sufficiently larger system, with which it can exchange energy. This
point of view provides several natural restrictions on the applicability of the geometric method itself. These
are often to do with constraints imposed by the system volume \cite{Rupp},\cite{RSSS},\cite{RuppAJP}. 
These constraints are listed here in Eq. (\ref{conditionsmain}). 

In black hole thermodynamics, the Riemannian geometric method is far more restricted than examples
involving conventional fluids, the prime reason being that one of key the assumptions in the latter, namely that
entropy is extensive, does not hold for black holes. Nevertheless, it has been common to use the geometric method purely as a 
mathematical tool prior to the advent of extended phase space thermodynamics \cite{Ray}, when the idea of
the thermodynamic black hole volume was somewhat obscure. However, with the recent understanding that a
variable cosmological constant can indeed be identified the black hole pressure and has a well defined
conjugate volume, it is important to ask how far one can proceed with geometric methods, given that
the restrictions due to the system volume given in Eq. (\ref{conditionsmain}) are now far more transparent. 

In this spirit, we have studied the geometric formalism for the four dimensional RN-AdS 
and Kerr-AdS black holes in extended phase space thermodynamics -- systems that have received considerable 
attention of late. We have shown that in a dimensionally consistent analysis, the scalar curvature 
is proportional to the thermodynamic volume of the black hole. In this sense, it is {\it not} universal, as
the thermodynamic volume of the RN-AdS (Kerr-AdS) black hole necessarily depends on its 
charge (angular momentum). However, the scalar curvature per unit volume is an universal quanitity,
much like fluid systems, as was pointed out in early studies of the geometry of fluid systems. 
Here, since the specific heat at constant volume vanishes, one has to more appropriately consider 
a normalized scalar curvature, following \cite{WeiLiuMannPRL}. 

We have considered three physical constraints on the normalized scalar curvature $R_{o}c_v$ in the geometry
of extended phase space thermodynamics, as summarized in Eq. (\ref{conditionsmain}): a) It must be larger
than the minimum volume allowed at a certain pressure, b) It must be larger than the specific volume, which sets
the lowest scale of the system and c) Away from criticality, it must be small compared to the overall volume of the system, 
for the Gaussian approximation on which the geometric formulation rests, to be valid. We find that all three might be
seriously challenged in extended phase space black hole thermodynamics. This points to the fact that the
application of geometric methods in this formalism as elaborated upon by several authors of late, 
has to be undertaken with great care and this is the main takeaway message from the results presented in this paper. 

\vskip0.2in
\noindent
{\bf Acknowledgments} : The work of T. S. is supported in part by Science and Engineering
Research Board (India) via Project No. EMR/2016/008037.

\end{document}